

\def\singlespace{\baselineskip=\normalbaselineskip}

\def\oneandahalfspace{\baselineskip=\normalbaselineskip
  \multiply\baselineskip by 3 \divide\baselineskip by 2}

\parskip=\medskipamount
\overfullrule=0pt
\raggedbottom
\def\normalparindent{24pt}
\nopagenumbers
\footline={\ifnum\pageno=1{\hfil}\else{\hfil\rm\folio\hfil}\fi}
\def\endpage{\vfill\eject}
\def\beginlinemode{\endmode\begingroup\parskip=0pt
                   \obeylines\def\\{\par}\def\endmode{\par\endgroup}}
\def\beginparmode{\endmode\begingroup \def\endmode{\par\endgroup}}
\let\endmode=\par
\def\raggedcenter{
                  \leftskip=2em plus 6em \rightskip=\leftskip
                  \parindent=0pt \parfillskip=0pt \spaceskip=.3333em
                  \xspaceskip=.5em\pretolerance=9999 \tolerance=9999
                  \hyphenpenalty=9999 \exhyphenpenalty=9999 }
\def\\{\cr}
\let\rawfootnote=\footnote\def\footnote#1#2{{\parindent=0pt\parskip=0pt
        \rawfootnote{#1}{#2\hfill\vrule height 0pt depth 6pt width 0pt}}}
\def\title{\null\vskip 3pt plus 0.2fill\beginlinemode\raggedcenter\bf}
\def\author{\vskip 3pt plus 0.2fill \beginlinemode\raggedcenter}
\def\affil{\vskip 3pt plus 0.1fill\beginlinemode\raggedcenter\it}
\def\abstract{\vskip 3pt plus 0.3fill \beginparmode{\noindent  ABSTRACT:~}  }
\def\endtitlepage{\endpage\body}
\def\body{\beginparmode\parindent=\normalparindent}
\def\head#1{\par\goodbreak{\immediate\write16{#1}
           {\noindent\bf #1}\par}\nobreak\nobreak}

\def\refto#1{$^{[#1]}$}
\def\ref#1{Ref.~#1}
\def\Ref#1{Ref.~#1}\def\cite#1{{#1}}\def\[#1]{[\cite{#1}]}

\def\(#1){(\call{#1})}
\def\call#1{{#1}}\def\taghead#1{{#1}}

\def\references{\head{REFERENCES}\beginparmode\frenchspacing\parskip=0pt}
\gdef\refis#1{\item{#1.\ }}
\gdef\journal#1,#2,#3,#4.{#1~{\bf #2}, #3 (#4)}
\def\endreferences{\body}
\def\endit{\endmode\vfill\supereject}\let\endpaper=\endit

\def\gsim{\mathrel{\raise.3ex\hbox{$>$\kern-.75em\lower1ex\hbox{$\sim$}}}}
\def\lsim{\mathrel{\raise.3ex\hbox{$<$\kern-.75em\lower1ex\hbox{$\sim$}}}}
\def\square{\kern1pt\vbox{\hrule height 0.6pt\hbox{\vrule width 0.6pt\hskip 3pt
   \vbox{\vskip 6pt}\hskip 3pt\vrule width 0.6pt}\hrule height 0.6pt}\kern1pt}
\def\sla{\raise.15ex\hbox{$/$}\kern-.72em}

\catcode`@=11
\newcount\r@fcount \r@fcount=0\newcount\r@fcurr
\immediate\newwrite\reffile\newif\ifr@ffile\r@ffilefalse
\def\w@rnwrite#1{\ifr@ffile\immediate\write\reffile{#1}\fi\message{#1}}
\def\writer@f#1>>{}
\def\referencefile{\r@ffiletrue\immediate\openout\reffile=\jobname.ref%
  \def\writer@f##1>>{\ifr@ffile\immediate\write\reffile%
    {\noexpand\refis{##1} = \csname r@fnum##1\endcsname = %
     \expandafter\expandafter\expandafter\strip@t\expandafter%
     \meaning\csname r@ftext\csname r@fnum##1\endcsname\endcsname}\fi}%
  \def\strip@t##1>>{}}

\def\citeall#1{\xdef#1##1{#1{\noexpand\cite{##1}}}}
\def\cite#1{\each@rg\citer@nge{#1}}
\def\each@rg#1#2{{\let\thecsname=#1\expandafter\first@rg#2,\end,}}
\def\first@rg#1,{\thecsname{#1}\apply@rg}
\def\apply@rg#1,{\ifx\end#1\let\next=\relax%
\else,\thecsname{#1}\let\next=\apply@rg\fi\next}%
\def\citer@nge#1{\citedor@nge#1-\end-}
\def\citer@ngeat#1\end-{#1}
\def\citedor@nge#1-#2-{\ifx\end#2\r@featspace#1
  \else\citel@@p{#1}{#2}\citer@ngeat\fi}
\def\citel@@p#1#2{\ifnum#1>#2{\errmessage{Reference range #1-#2\space is bad.}
    \errhelp{If you cite a series of references by the notation M-N, then M and
    N must be integers, and N must be greater than or equal to M.}}\else%
{\count0=#1\count1=#2\advance\count1 by1\relax\expandafter\r@fcite\the\count0,%
  \loop\advance\count0 by1\relax
    \ifnum\count0<\count1,\expandafter\r@fcite\the\count0,%
  \repeat}\fi}
\def\r@featspace#1#2 {\r@fcite#1#2,}    \def\r@fcite#1,{\ifuncit@d{#1}
    \expandafter\gdef\csname r@ftext\number\r@fcount\endcsname%
    {\message{Reference #1 to be supplied.}\writer@f#1>>#1 to be supplied.\par
     }\fi\csname r@fnum#1\endcsname}
\def\ifuncit@d#1{\expandafter\ifx\csname r@fnum#1\endcsname\relax%
\global\advance\r@fcount by1%
\expandafter\xdef\csname r@fnum#1\endcsname{\number\r@fcount}}
\let\r@fis=\refis   \def\refis#1#2#3\par{\ifuncit@d{#1}%
    \w@rnwrite{Reference #1=\number\r@fcount\space is not cited up to now.}\fi%
  \expandafter\gdef\csname r@ftext\csname r@fnum#1\endcsname\endcsname%
  {\writer@f#1>>#2#3\par}}
\def\r@ferr{\endreferences\errmessage{I was expecting to see
\noexpand\endreferences before now;  I have inserted it here.}}
\let\r@ferences=\references
\def\references{\r@ferences\def\endmode{\r@ferr\par\endgroup}}
\let\endr@ferences=\endreferences
\def\endreferences{\r@fcurr=0{\loop\ifnum\r@fcurr<\r@fcount
    \advance\r@fcurr by 1\relax\expandafter\r@fis\expandafter{\number\r@fcurr}%
    \csname r@ftext\number\r@fcurr\endcsname%
  \repeat}\gdef\r@ferr{}\endr@ferences}
\let\r@fend=\endpaper\gdef\endpaper{\ifr@ffile
\immediate\write16{Cross References written on []\jobname.REF.}\fi\r@fend}
\catcode`@=12
\citeall\refto\citeall\ref\citeall\Ref
\catcode`@=11
\newcount\tagnumber\tagnumber=0
\immediate\newwrite\eqnfile\newif\if@qnfile\@qnfilefalse
\def\write@qn#1{}\def\writenew@qn#1{}
\def\w@rnwrite#1{\write@qn{#1}\message{#1}}
\def\@rrwrite#1{\write@qn{#1}\errmessage{#1}}
\def\taghead#1{\gdef\t@ghead{#1}\global\tagnumber=0}
\def\t@ghead{}\expandafter\def\csname @qnnum-3\endcsname
  {{\t@ghead\advance\tagnumber by -3\relax\number\tagnumber}}
\expandafter\def\csname @qnnum-2\endcsname
  {{\t@ghead\advance\tagnumber by -2\relax\number\tagnumber}}
\expandafter\def\csname @qnnum-1\endcsname
  {{\t@ghead\advance\tagnumber by -1\relax\number\tagnumber}}
\expandafter\def\csname @qnnum0\endcsname
  {\t@ghead\number\tagnumber}
\expandafter\def\csname @qnnum+1\endcsname
  {{\t@ghead\advance\tagnumber by 1\relax\number\tagnumber}}
\expandafter\def\csname @qnnum+2\endcsname
  {{\t@ghead\advance\tagnumber by 2\relax\number\tagnumber}}
\expandafter\def\csname @qnnum+3\endcsname
  {{\t@ghead\advance\tagnumber by 3\relax\number\tagnumber}}
\def\equationfile{\@qnfiletrue\immediate\openout\eqnfile=\jobname.eqn%
  \def\write@qn##1{\if@qnfile\immediate\write\eqnfile{##1}\fi}
  \def\writenew@qn##1{\if@qnfile\immediate\write\eqnfile
    {\noexpand\tag{##1} = (\t@ghead\number\tagnumber)}\fi}}
\def\callall#1{\xdef#1##1{#1{\noexpand\call{##1}}}}
\def\call#1{\each@rg\callr@nge{#1}}
\def\each@rg#1#2{{\let\thecsname=#1\expandafter\first@rg#2,\end,}}
\def\first@rg#1,{\thecsname{#1}\apply@rg}
\def\apply@rg#1,{\ifx\end#1\let\next=\relax%
\else,\thecsname{#1}\let\next=\apply@rg\fi\next}
\def\callr@nge#1{\calldor@nge#1-\end-}\def\callr@ngeat#1\end-{#1}
\def\calldor@nge#1-#2-{\ifx\end#2\@qneatspace#1 %
  \else\calll@@p{#1}{#2}\callr@ngeat\fi}
\def\calll@@p#1#2{\ifnum#1>#2{\@rrwrite{Equation range #1-#2\space is bad.}
\errhelp{If you call a series of equations by the notation M-N, then M and
N must be integers, and N must be greater than or equal to M.}}\else%
{\count0=#1\count1=#2\advance\count1 by1\relax\expandafter\@qncall\the\count0,%
  \loop\advance\count0 by1\relax%
    \ifnum\count0<\count1,\expandafter\@qncall\the\count0,  \repeat}\fi}
\def\@qneatspace#1#2 {\@qncall#1#2,}
\def\@qncall#1,{\ifunc@lled{#1}{\def\next{#1}\ifx\next\empty\else
  \w@rnwrite{Equation number \noexpand\(>>#1<<) has not been defined yet.}
  >>#1<<\fi}\else\csname @qnnum#1\endcsname\fi}
\let\eqnono=\eqno\def\eqno(#1){\tag#1}\def\tag#1$${\eqnono(\displayt@g#1 )$$}
\def\aligntag#1\endaligntag  $${\gdef\tag##1\\{&(##1 )\cr}\eqalignno{#1\\}$$
  \gdef\tag##1$${\eqnono(\displayt@g##1 )$$}}
\def\eqalignno#1{\displ@y \tabskip\centering
  \halign to\displaywidth{\hfil$\displaystyle{##}$\tabskip\z@skip
    &$\displaystyle{{}##}$\hfil\tabskip\centering
    &\llap{$\displayt@gpar##$}\tabskip\z@skip\crcr
    #1\crcr}}
\def\displayt@gpar(#1){(\displayt@g#1 )}
\def\displayt@g#1 {\rm\ifunc@lled{#1}\global\advance\tagnumber by1
        {\def\next{#1}\ifx\next\empty\else\expandafter
        \xdef\csname @qnnum#1\endcsname{\t@ghead\number\tagnumber}\fi}%
  \writenew@qn{#1}\t@ghead\number\tagnumber\else
        {\edef\next{\t@ghead\number\tagnumber}%
        \expandafter\ifx\csname @qnnum#1\endcsname\next\else
        \w@rnwrite{Equation \noexpand\tag{#1} is a duplicate number.}\fi}%
  \csname @qnnum#1\endcsname\fi}
\def\ifunc@lled#1{\expandafter\ifx\csname @qnnum#1\endcsname\relax}
\let\@qnend=\end\gdef\end{\if@qnfile
\immediate\write16{Equation numbers written on []\jobname.EQN.}\fi\@qnend}
\catcode`@=12

\magnification=1200
\oneandahalfspace

\footnote{}{\singlespace $^*$\  Essay submitted to the Gravity
Research Foundation}
\title MIDISUPERSPACE-INDUCED CORRECTIONS TO THE WHEELER DE WITT EQUATION  $^*$
\author Francisco D. Mazzitelli
\affil
International Centre for Theoretical Physics
P.O. Box 586 - 34100 Trieste - Italy
\abstract
We consider the midisuperspace of four dimensional spherically
symmetric  metrics and the Kantowski-Sachs minisuperspace
contained in it.  We discuss the quantization of the midisuperspace using
the fact that the dimensionally reduced Einstein
Hilbert action  becomes  a scalar-tensor theory of gravity in
two dimensions. We show that the covariant regularization procedure in the
midisuperspace induces modifications into the minisuperspace
Wheeler DeWitt equation.

\bigskip

\bigskip

\centerline {{\bf March 1992}}

\endtitlepage

\bigskip

Proposed by DeWitt and Misner $^1$ more than twenty years ago,
the minisuperspace approximation has been almost the only practical
way  of dealing with the difficulties of quantizing  General Relativity.

Although it was clear  from the
very beginning that the minisuperspace approximation could
be considered only as a crude model of quantum gravity,
the question of the validity of the approximation has
been addressed quantitatively only in the last  years.
Kuchar and Ryan $^2$ have investigated a hierarchy of minisuperspace
models trying to find a criteria to decide when the quantum
minisuperspace results are meaningful. Alternatively,
Hu and collaborators $^3$,
have developed a formalism to obtain an effective Wheeler DeWitt
equation in the minisuperspace which takes into account the
backreaction of the ignored modes. In any case, and because a
complete theory of quantum gravity is still lacking, it seems that
the only way to learn what are we missing when we use the
minisuperspace approximation is to embed the minisuperspace into
a larger one, analyze the enlarged theory and compare
the results with the ones obtained in the original minisuperspace.

In the present work, we  will follow this procedure. We will study
the midisuperspace of spherically symmetric four dimensional metrics and the
Kantowski-Sachs minisuperspace contained in it. We will
see how the gauge fixing and covariant regularization procedures in the
midisuperspace induce
important modifications into the minisuperspace-Wheeler DeWitt equation.

Let us consider  the spherically symmetric metrics
$$ds^2= g_{ab}(x^0,x^1)dx^adx^b+2G\Psi (x^0,x^1)(d\theta ^2+\sin ^2\theta
d\varphi^2)\eqno (1)$$
where $x^0=t, x^1=r, \theta $ and $\varphi$ are coordinates
adapted to the spherical symmetry. The
Newton constant G is inserted in order to make $\Psi$
dimensionless. We will assume that the radial coordinate is compact.

The four dimensional Einstein-Hilbert action
$$ S={1\over 16\pi G}\int d^4x\sqrt{g^{(4)}}\left (R^{(4)}+\Lambda\right
)\eqno(2)$$
reads, after inserting the ansatz (1),
$$S={1\over 2}\int d^2x\sqrt g\left (R\Psi+{1\over G}+\Lambda \Psi
+{g^{ab}\over 2\Psi}\Psi_{,a}\Psi_{,b}\right)\quad\quad .\eqno (3)$$
It can be explicitly checked that the Euler-Lagrange equations associated
with the action (2) coincide with the Einstein equations for
spherically symmetric metrics.$^4$

We can interpret the dimensionally reduced action (3) as a scalar-tensor
theory of gravity in two dimensions. The action
can be further simplified by adopting the conformal gauge for
the two dimensional metric $g_{ab}$, that is $g_{ab}=\hat g_{ab}\exp\lambda$.
In this gauge we have
$$S=\int d^2x\sqrt{\hat g}\left ({1\over 2}G_{ij}(X)\hat g^{ab}\partial_aX^i
\partial_bX^j+{1\over 2}\hat RD(X)+T(X)\right )\eqno (4)$$
where
$$X^i=\pmatrix{\lambda\cr \Psi\cr}$$
$$G_{ij}(X)={1\over 2}\pmatrix{0&1\cr 1&\Psi^{-1}\cr}$$
$$D(X)=\Psi$$
$$T(X)={1\over 2}({1\over G}+\Lambda \Psi)\exp\lambda$$
We recognize that Eq. (4) corresponds to a two dimensional non linear
sigma-model
with target space metric $G_{ij}(X)$, dilaton field $D(X)$ and tachyon field
$T(X)$.

As we have mentioned, the spherically symmetric midisuperspace in Eq. (1)
contains as a particular case the Kantowski-Sachs minisuperspace
$$ds^2=a^2(t)\left( -dt^2+dr^2\right)+b^2(t)(d\theta ^2+\sin ^2\theta
d\varphi^2)\quad\quad ,\eqno (5)$$
which describes a $S^1\times S^2$
universe. The Wheeler DeWitt equation (WDWE) associated with this
minisuperspace
can be obtained from (4) assuming that the fields $X^i$ are only functions of
$t$ and setting
$$\hat g_{ab}=\pmatrix{-N^2(t)&0\cr 0&1\cr}\quad\quad .$$
Here $N$ is the lapse function.
After standard manipulations we find,
$$\left [-{1\over 2}\nabla^2+T(X)\right ]\psi (X)=0\quad\quad ,\eqno (6)$$
where $\nabla^2$ is the Laplacian of the supermetric $G_{ij}$. We have
chosen a covariant factor ordering in minisuperspace.

Now comes the crucial point of this essay. Before assuming the spatial
homogeneity of $X^i$, the quantum theory associated to the midisuperspace
(1) is a two dimensional {\it quantum field theory}. The quantum fields
are $\lambda (r,t)$ and $\Psi (r,t)$. As in any field
theory, a regularization is needed, and in our case this regularization must
be covariant with respect to the full metric $g_{ab}=\hat g_{ab}\exp \lambda$.
As a consequence, the theory is complicated by the fact  there is an additional
$\lambda$-dependence
in the path-integral measure $[d\lambda d\Psi]_{\hat g\exp\lambda}$.
This is a very
well known problem in  $2d$ gravity coupled to conformally
invariant matter. There,  it was conjectured by David, Distler and Kawai $^5$
(DDK),
 that the $\lambda$-dependence of the measure can be taken
into account  as follows
$$\int [d\lambda]_{\hat g\exp\lambda}\exp{(-S_{L})}=\int [d\lambda]_{\hat
g}\exp{(-\bar S_{L})}$$
where $S_L$ is the Liouville action and $\bar S_L$ is an
action of the same form but with different coupling constants. The new
coefficients in $\bar S_{L}$ are fixed by requiring the
quantum theory to be invariant under the transformation
$$\hat g_{ab}\rightarrow\hat g_{ab}\exp\tau
(x)\quad\quad\quad\lambda\rightarrow\lambda
-\tau (x)\quad\quad .\eqno (7)$$
This is an obvious  requirement since the {\it full} metric  $g_{ab}=\hat
g_{ab}\exp \lambda$ is left
unchanged and it is only $g_{ab}$ that enters in both  the original theory
and the regulator.

In our case, the classical action is a non linear sigma-model in two
dimensions.
Power counting implies that, at the quantum level, the theory is renormalizable
in a generalized  sense. One must  allow  for a change in the form of
the metric, the dilaton and the tachyon fields.
A DDK-like argument then suggests $^6$ that the $\lambda$
dependence of the measure can be taken into account by
considering, instead of the classical action (4), an action $\bar S$  of the
same form but with general couplings $\bar G_{ij}(X)$, $\bar D(X)$ and
$\bar T(X)$. As before, the couplings must be such that the `split'
transformation Eq. (7) is an exact quantum symmetry.
As the new measure  $[d\lambda d\Psi]_{\hat g}$ is translationally invariant,
transformation (7) implies Weyl invariance with respect to $\hat g_{ab}$.
As a consequence,
the beta functions of the couplings should vanish, that is $^7$
$$\eqalign{\beta_{\bar G}&=0=\bar R_{ij}+2\bar\nabla_i\bar\nabla_j\bar D\cr
\beta_{\bar D}&=0=-c-{1\over 2}\bar\nabla^2\bar D+\bar\nabla\bar D .
\bar\nabla\bar D\cr
\beta_{\bar T}&=0=-{1\over 2}\bar\nabla^2\bar T+\bar\nabla\bar T
.\bar\nabla\bar D-2\bar T}\eqno (8)$$
where we omited quadratic terms in the tachyon field.
The constant $c$ in the second equation is the Weyl anomaly coefficient, given
by
$c={1\over 6}(26-d)$, where $d$ is the dimension of the target space.
In our case we have $c=4$.
We stress that
the Ricci tensor and the covariant derivatives in Eqs. (8) are
formed with the target space metric $\bar G_{ij}$.

In principle one can go ahead and quantize the theory based on the
new action using the $\lambda$-independent measure $[d\lambda d\Psi]_{\hat g}$.
The physical states can be obtained from the BRST cohomology.
This is beyond the scope of this essay. We will
simply implement the minisuperspace approximation in the
`effective' action $\bar S$, that is, after the $\lambda$ dependence
of the measure has been removed $^8$. The physical states must
then satisfy the modified WDWE
$$\left [-{1\over 2}\bar\nabla^2+\bar T(X)\right ]\psi (X)=0\eqno (9)$$
where $\bar\nabla^2$ is the Laplacian of  $\bar G_{ij}$.
Both $\bar G_{ij}$ and $\bar T$ must satisfy Eqs. (8). This is our main result.

It is easy to find  solutions to the equations $\beta_{\bar G}=0$ and
$\beta_{\bar D}=0$
which coincide with the classical $G_{ij}$ and $D$ when $c=0$. In fact, one
can show that
$$\bar G_{ij}(X)={1\over 2}\pmatrix{c/2&1\cr 1&\Psi^{-1}\cr}$$
$$\bar D(X)=\Psi+{c\over 2}\lambda\eqno (10)$$
satisfy both equations.
With these $\bar G_{ij}$ and $\bar D$  one can solve the
tachyon equation $\beta_{\bar T}=0$ using the classical $T$ as
a boundary condition.
We conclude that the minisuperspace WDWE has been modified in two
ways. On the one hand, the Weyl anomaly $c\neq 0$ modifies the structure
of the differential operator. As $\bar G_{\lambda\lambda}\neq 0$,
the operator $\bar\nabla^2$ will contain a term with second derivatives
with respect to $\Psi$. This term is absent in $\nabla^2$. On the other
hand, the new superpotential $\bar T$ will differ from the original
one.

Summarizing, we arrived at the following picture.
In the minisuperspace approximation the Wheeler DeWitt equation (6) is
constructed
with the classical supermetric $G_{ij}$ and superpotential $T$.
After taking into account that the midisuperspace quantum field theory
must be covariantly regularized,
the classical supermetric and
superpotential
must be replaced by the quantum couplings $\bar G_{ij}$ and $\bar T$.
We expect this to be a general feature, not restricted to the
Kantowski-Sachs example considered here. That is, similar modifications
should take place in other minisuperspace examples.

For simplicity we considered a minisuperspace approximation to the
modified action $\bar S$. A more complete treatment of the problem should
 include the backreaction  of the neglected inhomogeneous modes on
the minisuperspace variables $^3$. We expect this backreaction to produce
additional non local terms in the modified Wheeler DeWitt equation.

\bigskip

\noindent
{\bf Acknowledgements:} I would like to thank N. Mohammedi for his
helpful comments
and  Prof. Abdus Salam, IAEA and UNESCO for financial support.

\vfill\eject
\noindent
{\bf REFERENCES}
\parindent=0pt

\item {1.} B. DeWitt, Phys. Rev. 160, 1113 (1967); C.W. Misner,
Phys. Rev. 186, 1319 (1969)

\item {2.} K. Kuchar and M.P. Ryan, Phys. Rev. D40, 3982 (1989)

\item {3.} B.L. Hu, Lectures delivered at the Seventh Latin American
International Symposium on General Relativity and Gravitation, Mexico, 1990,
and references therein.

\item {4.} P. Thomi, B. Isaak and P. Hajicek, Phys. Rev. D30, 1169 (1984)

\item {5.} F. David, Mod. Phys. Lett A3, 1651 (1988); J. Distler and
H. Kawai, Nucl. Phys. B321, 509 (1989)

\item {6.} A.A. Tseytlin, Int. J. Mod. Phys. A5, 1833 (1990);
J.G. Russo and A.A. Tseytlin, preprint SU-ITP 92-2 (1992);
A. Cooper, L. Susskind and L. Thorlacius, Nucl.Phys. B363, 132 (1991)

\item {7.} C.G. Callan, D. Friedan, E. Martinec and M.J. Perry,
Nucl. Phys. B262, 593 (1985)

\item {8.}  A similar approximation in $2d$ gravity has been done in
J. Polchinski, Nucl. Phys. B324, 123 (1989)

\end